\documentclass{elsart5p}
\usepackage{graphicx}
\usepackage{lineno}
\begin{document}
\begin{frontmatter}
\title{Development of a resonant laser ionization gas cell for high-energy, short-lived nuclei}
\author[label1]{T.~Sonoda\corauthref{cor}},
\corauth[cor]{Corresponding author. Tel.: +81-48-467-8640; fax: +81-48-462-7563.}
\ead{tetsu@riken.jp}
\author[label1]{M.~Wada},
\author[label2,label1]{H.~Tomita},
\author[label2,label1]{C.~Sakamoto},
\author[label2,label1]{T.~Takatsuka},
\author[label3,label1]{T.~Furukawa},
\author[label4,label1]{H.~Iimura},
\author[label3,label1]{Y.~Ito},
\author[label1]{T.~Kubo},
\author[label1]{Y.~Matsuo},
\author[label5,label1]{H.~Mita},
\author[label1]{S.~Naimi},
\author[label5,label1]{S.~Nakamura},
\author[label2,label1]{T.~Noto},
\author[label5,label1]{P.~Schury},
\author[label6]{T.~Shinozuka},
\author[label6]{T.~Wakui},
\author[label7]{H.~Miyatake},
\author[label7]{S.~Jeong},
\author[label7]{H.~Ishiyama},
\author[label7]{Y.X.~Watanabe},
\author[label7]{Y.~Hirayama},
\author[label8,label1]{K.~Okada},
\author[label9,label1]{A.~Takamine},
\\
\address[label1]{RIKEN Nishina Center for Accelerator-Based Science, 2-1 Hirosawa, Wako, Saitama 351-0198, Japan}
\address[label2]{Faculty of Engineering, Nagoya University, Nagoya 464-8603, Japan}
\address[label3]{Department of Physics,Tokyo Metropolitan University, Tokyo 116-8551, Japan}
\address[label4]{Japan Atomic Energy Agency (JAEA), Tokaimura 319-1100, Japan}
\address[label5]{Department of Physics, Tsukuba University, Tsukuba 305-8577, Japan}
\address[label6]{Cyclotron and Radioisotope Center, Tohoku University, Sendai 980-8578, Japan}
\address[label7]{High Energy Accelerator Research Organization (KEK) 305-0801, Japan}
\address[label8]{Department of Physics, Sophia University, 7-1 Kioicho, Chiyoda, Tokyo 102-8554, Japan}
\address[label9]{Department of Physics, Aoyama Gakuin University, 4-4-25 Shibuya, Shibuya-ku, Tokyo 150-8366, Japan}
\begin{abstract}
A new laser ion source configuration based on resonant photoionization in a gas cell has been developed at RIBF RIKEN. This system is intended for the future PArasitic RI-beam production by Laser Ion-Source (PALIS) project which will be installed at RIKEN's fragment separator, BigRIPS.  A novel implementation of differential pumping, in combination with a sextupole ion beam guide (SPIG), has been developed. A few small scroll pumps create a pressure difference from 1000 hPa $\sim$ 10$^{-3}$ Pa within a geometry drastically miniaturized compared to conventional systems. 
This system can utilize a large exit hole for fast evacuation times, minimizing the decay loss for short-lived nuclei during extraction from a buffer gas cell, while sufficient gas cell pressure is maintained for stopping high energy RI-beams. 
In spite of the motion in a dense pressure gradient, the photo-ionized ions inside the gas cell are ejected with an assisting force gas jet and successfully transported to a high-vacuum region via SPIG followed by a quadrupole mass separator.
Observed behaviors agree with the results of gas flow and Monte Carlo simulations.
\end{abstract}


\end{frontmatter}

\section{Introduction}
Radioactive ion beam (RIB) facilities based on the in-flight production technique provide a wide variety of exotic nuclei without restrictions on lifetimes or chemical properties. An essential requirement for present and future RIB facilities is to transform this high-energy beam into a low-energy, low-emmitance beam. Such low-energy beams open up opportunities to study ground state properties of exotic nuclei by experimental techniques such as laser spectroscopy and ion trapping. At RIKEN,  a universal slow RI-beam facility, SLOWRI, based on a gas catcher cell with an RF-carpet ion guide \cite{w1}\cite{rk2}, was assigned as one of the principal facilities of RIBF. A novel method, named PALIS (PArasitic RI-beam production by Laser Ion-Source) \cite{rk1} was also approved for the construction, to expand the usability and reduce experimental costs by utilizing unused RI-beams produced by projectile fragmentation or in-flight fission. Using this scheme,  it will be possible to perform low-energy RI-beam experiments alongside every on-line BigRIPS experiment.  

\begin{figure}[h]
\begin{center}
 \includegraphics*[width=8cm]{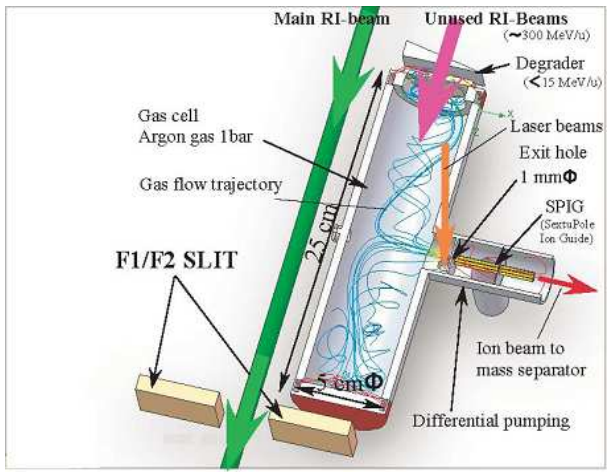}
\end{center}
\caption{Schematic of the laser ionization gas catcher setup. Without disturbing the main beam, an unwanted portion of the beam is stopped in the argon gas cell.  Neutralized atoms are transported by gas flow towards the exit and re-ionized by resonant laser radiations. Thusly produced low-energy RI-beams can be utilized for slow-RI beam experiments.}
\label{fig1}
\end{figure}

At RIKEN RIBF, the RI-beams of highly exotic nuclei are available with the highest intensity in the world. However, the usability is restricted to short periods of beam time due to high demand and limited yearly operating hours due to the electrical cost for accelerator operation. To bring about the most effective utilization of such high-performance facility, parasitic production of unused RI-beams would be valuable.  
In-flight fission and fragmentation produces a beam which is a mixture of thousands of isotopes. A fragment separator selects one specific RI-beam by removing the vast majority of these isotopes.  The removed isotopes still include many rare nuclei of interest for nuclear studies; wastefully, they are simply thrown away. Our method is to save these rare isotopes before their removal in a beam purification slit. By installing a gas catcher cell in the vicinity of the slit at the F1 or  F2 focal plane in BigRIPS \cite{bg1}, the RI which would have been removed by the slit can be collected and salvaged as a low-energy RI-beam. A schematic view of this setup is shown in Fig.~\ref{fig1}.
Due to space and accessibility constraints, a big gas cell such as the ones typically used in RIB facilities \cite{w1}\cite{gs1}\cite{gs2} is not possible.  Instead, we must use a compact cell with a simpler mechanism. Such a compact gas catcher cell necessitates higher pressure in order to maintain enough stopping power for the high energy RI-beams. Therefore, we will use a laser ionization gas cell \cite{ls0}-\cite{shd} which can be a few hundred cm$^{3}$ in volume with 1 bar argon gas. The thermalized ions are quickly neutralized in high-pressure argon gas and transported by  gas flow toward the exit of the cell,  where they are selectively re-ionized by resonant laser radiations in the vicinity of the exit. They can then be further purified by an electro-magnetic mass separator and transported to the low-energy experimental room. In this way, the parasitic low-energy RI-beams could be delivered whenever BigRIPS experiments are in operation. 

In order to provide a high stopping efficiency for high-energy RI-beams, the pressure in the gas cell must be sufficient to ensure the stopping range is shorter than the finite length of the gas cell. Additionally, the transport of neutral atoms by gas flow constrains the volume of the gas cell due to diffusion loss.  Thus, it is optimal to use the highest possible pressure to allow the smallest possible gas cell. Additionally, a fast evacuation time is necessary to avoid decay losses for short-lived nuclei during transportation inside the gas cell. This can be accomplished by enlarging the size of the exit hole. For dealing with such high gas throughput, while keeping a high-vacuum extraction beam line, large roots pumps with pumping speed of the order of 10$^{3}$ m$^{3}$/h are typically used in a differential pumping system. The radiofrequency sextupole ion beam guide (SPIG) \cite{sp1}-\cite{sp4} can moderate the pumping load by decreasing the conductance between  the gas cell and the extraction chamber. Even using such high-throughput pumps and SPIG, however, there are  limits to the allowable pressure and consequently the number of feasible RIs extracted from the gas cell due to a lack of stopping efficiency and a relatively slow evacuation time.
In order to address such inherent limitations, while also miniaturizing the entire system to fit within the highly constrained space limitations of the PALIS installation, we have developed a new idea which is the stepwise differential pumping method by small pumping capacities.  
This enables use of an even larger exit hole, more than 1 mm in a diameter, with the use of pressures up to one atmosphere argon in the gas cell.  For such a pressure and exit hole, RI-beams with energies up to 10 MeV per nucleon will have a stopping range within the length of the gas cell (25 cm), while the evacuation time of the cell (gas cell volume $\sim$500 cm$^{3}$) will allow a half-life of 100 ms to be extracted with 20\% efficiency. 

To provide proof of principle for this new gas cell based laser ionization system, a prototype gas cell and a beam extraction system has been built for off-line experiments. 
The differential pumping capability of this new system has been verified; a pressure difference from 1000 hPa argon in the gas cell down to 10$^{-3}$ Pa at the quadrupole mass filter has been achieved, while using a 1 mm diameter gas cell exit hole. This is the first result obtained in this mode: resonant laser ionization inside the gas cell, along with ion extraction from the gas cell and transport to high-vacuum. This has been experimentally confirmed off-line for several stable elements produced by filament evaporation or ablation by YAG laser.
Resonant ionization has been performed using a two-step excitation scheme with an excimer-dye-laser combination. 
The extraction time profiles of the gas cell and the SPIG have been investigated and also examined by a Monte Carlo simulation combined with a gas flow calculation. The experimental results show that ions can be transported from a high pressure to a high-vacuum region via a long SPIG (253 mm) with an assisting force gas jet.
This technique will relieve the inherent restrictions of conventional Ion Guide Isotope Separator On-Line (IGISOL)  technique \cite{jy2} based gas catcher cells, providing faster extraction times and improved stopping efficiency.  

\section{Prototype gas cell and a beam extraction system}

\begin{figure}[h]
\begin{center}
 \includegraphics*[width= 9.3 cm,height=6.3 cm]{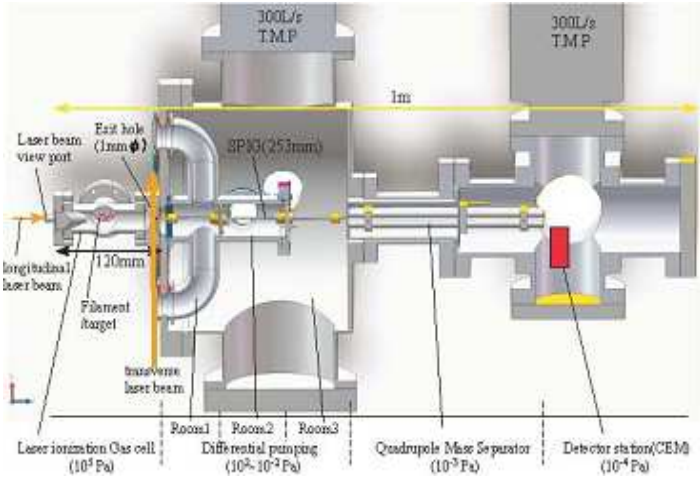}
\end{center}
\caption{Cut-away view of the prototype gas cell and the differential pumping system for PArasitic RI-beam production by Laser Ion-Source (PALIS). }
\label{fig2}
\end{figure}

A prototype gas cell and a beam extraction system has been developed. A conceptual sketch is shown in Fig.~\ref{fig2}. The system is composed of four parts: a laser ionization gas cell, a differential pumping system, a quadrupole mass separator and a detector station comprising a Channel Electron-Multiplier (CEM).
In this system, differential pumping -- from the high pressure gas cell to the ion detector in a high-vacuum region -- is achieved in just 1 m by applying our novel differential pumping method without using large roots pumps. 

\subsection{The gas cell}
\begin{figure}[h]
\begin{center}
 \includegraphics*[width= 7.3 cm,height=6.3 cm]{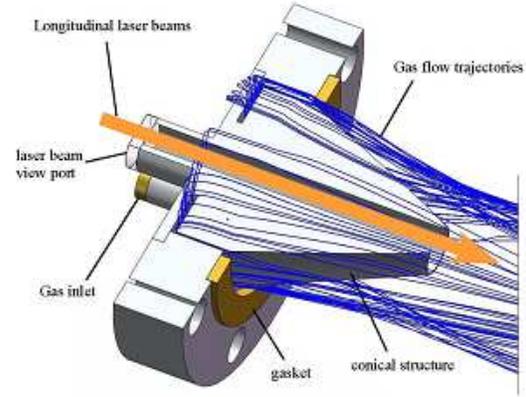}
\end{center}
\caption{The shape of the gas inlet was carefully fabricated by referencing flow simulation results. The gas is isotropically spread in the gas cell.}
\label{fig3}
\end{figure}

 In the present off-line setup, a gas cell consisting of a simple cross chamber with dimension of 120 $\times$ $\phi$ 70 mm was used. It has an exit hole of 1 mm diameter and two feedthroughs to provide an electric current to a filament. In order to create a laminar gas flow inside the cell, the shape of the gas inlet was carefully fabricated by referencing flow simulation results. The gas inlet structure is shown in Fig.~\ref{fig3}. The buffer gas is introduced via an inlet pipe of 6.35 mm diameter and spread with the conical structure.
 Laser beams can be introduced into the ionization region inside the gas cell via a quartz view port. Two alternative laser paths, transverse and longitudinal to the beam extraction axis, exist for ionization inside the gas cell. In the present experiment, however, only the transverse path was used.
In the off-line setup, there are two methods available for producing atomic vapor inside the gas cell: evaporation of a filament or ablation by YAG laser.  For laser ablation, the filament was replaced by a target which was irradiated by a YAG laser via a view port in the flange opposite the target. 
The gas handling system was specially designed to remove any impurity or contaminant from the argon buffer gas.
Electro-polished stainless steel tubes and metal sealed valves were used. The gas cell and gas line were heated to 200 $^\circ$C and pumped by a turbo molecular pump backed by a scroll pump. The argon gas was regulated using a needle valve and injected into the inlet of the gas cell via a purifier (SAES Pure Gas, Inc FT400-902).

\subsection{The differential pumping system}

\begin{figure}[ht]
\begin{center}
 \includegraphics*[width= 9.3 cm,height=6.3 cm]{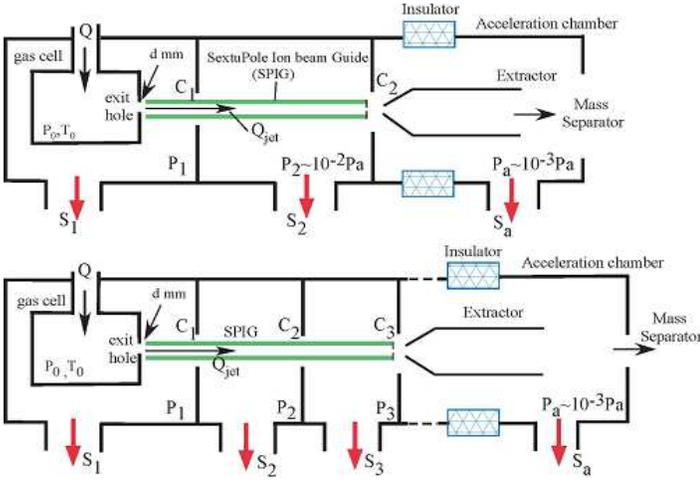}
\end{center}
\caption{Typical IGISOL layouts of the gas cell and its connection to the extraction chamber of the isotope separator (top) and conceptual layout  newly proposed differential pumping system connecting the gas cell to the mass separator (bottom).}
\label{fig4}
\end{figure}

The differential pumping system is key to realize the coupling of gas cell and isotope separator. Historically, it has been used in the IGISOL technique \cite{jy2}, where the pumping capability is a critical parameter to determine the operational pressure for buffer stopping gas and the size of the gas cell exit hole. High pressures required for stopping energetic RI-ions efficiently and a large exit hole for fast evacuation to avoid the decay losses in short-lived RIs make for a severe pumping load in the differential pumping system.  A typical IGISOL differential pumping system is shown in Fig.~\ref{fig4} (top).

When a compressible viscous gas flow passes through an orifice separating different pressures regions, the gas flow velocity becomes supersonic at a certain pressure difference. Then the flow rate $Q$ becomes independent of the pressure in the expansion area. The expression for $Q$ in this case is given by \cite{vceq}

\begin{equation}
Q=AP_{\rm{0}}\biggl(\frac{2}{\gamma+1}\biggl)^{\frac{1}{\gamma-1}}\biggl(\frac{R_{\rm{0}}T_{\rm{0}}}{M}\frac{2\gamma}{\gamma+1}\biggl)^{\frac{1}{2}} ,
\label{eqn:a}
\end{equation}

where $A$ is the cross section of the exit hole (cm$^{2}$), $P$$_{0}$ the pressure in the gas cell, $\gamma$ the
ratio of specific heats $C$$_{p}$/$C$$_{v}$, $R$$_{0}$ the gas constant, $M$ the atomic mass of the gas and $T$$_{0}$ the gas temperature.  

Additionally, some relationships relating flow rate to pumping speed and conductance can be described in this system as follows :

\begin{equation}
Q=Q_{\rm{jet}} + (P_{1}-P_{2})C_{1} + P_{1}S_{1} ,
\label{eqn:b}
\end{equation}

\begin{equation}
Q_{\rm{jet}} + (P_{1}-P_{2})C_{1} =(P_{2}-P_{a})C_{2} + P_{2}S_{2} ,
\label{eqn:c}
\end{equation}

\begin{equation}
(P_{2}-P_{a})C_{2} = P_{a}S_{a} ,
\label{eqn:d}
\end{equation}

where $Q$$_{\rm{jet}}$ is the fraction of the flow rate comprised by the jet component which is directed into the SPIG.  $P$$_{1}$, $P$$_{2}$ and $P$$_{a}$ are the pressures of the individual rooms: $P$$_{1}$ is the first room, containing the gas cell in the housing chamber on the beam line from accelerator;  $P$$_{2}$ is the middle room in the differential pumping region; and $P$$_{a}$ is the third room, labelled as acceleration chamber in Fig. \ref{fig4}. $C$$_{1}$ and $C$$_{2}$ are the  conductances from the orifices connected between the first to the middle and the middle to the third room.  $S$$_{1}$, $S$$_{2}$ and $S$$_{a}$ are the effective pumping speeds connected with individual room. 

 For proper ion acceleration without discharge in the separator, $P$$_{a}$ should be on the order of 10$^{-3}$ Pa or less.  Therefore, it is necessary for $P$$_{2}$ be on the order of 10$^{-2}$ Pa. 
This can be easily deduced from eq. (\ref{eqn:d}) by assuming the typical conductance of extractor electrode to be $C$$_{2}$ = 10$^{-1}$ m$^{3}$/s and standard pumping speed (a few m$^{3}$/s) of turbo molecular or diffusion pumps for $S$$_{2}$ and $S$$_{a}$. Consequently the last term in the right-hand  side of eq. (\ref{eqn:b}), $P$$_{1}$$S$$_{1}$, should be nearly equal to the total flow rate $Q$. This means that most of the gas flow emitted from the gas cell is evacuated by the first pump $S$$_{1}$,resulting in a severe pumping load. 
For instance if we take the following parameters: $S$$_{2}$ = 2.0 m$^{3}$/s, $S$$_{a}$ = 1.0 m$^{3}$/s, $P$$_{1}$ = 1.0 Pa, $P$$_{2}$ =  1.0 $\times$ 10$^{-2}$ Pa and $P$$_{3}$ =  1.0 $\times$ 10$^{-3}$ Pa, then pumping speed $S$$_{1}$ must be nearly 6 $\times$ 10$^{3}$ m$^{3}$/h in order to use exit hole diameter of 0.5 mm while keeping a gas cell pressure of a few hundred hPa argon.

In the present idea, however, the differential pumping rooms up to the acceleration chamber are divided into more than three separately pumped volumes, as shown in Fig.~\ref{fig4} (bottom). Instead of immediately removing the buffer gas  by a high-throughput pump, the pressure is stepwise decreased across several rooms by means of small pumps. 
Accordingly, the total flow rate Q can be also written as:

\begin{equation}
Q = \sum_{i=1}^{R} P_{i}S_{i} + P_{a}S_{a} .
\label{eqn:f}
\end{equation}

where $R$ is the number of differential pumping rooms.\\
Thus the majority of the flow rate $Q$  is divided by a number of evacuation rooms to achieve a high-vacuum. 
While the pumping speed $S$$_{a}$ required to achieve high-vacuum in the acceleration chamber still needs high-throughput turbo molecular pumps, $S$$_{i}$ can be achieved with small pumps, resulting in a very compact differential pumping system.

Based on this idea, we developed a differential pumping system as shown in Fig.~\ref{fig2}. There are three differential pumping rooms before high vacuum. The first and the second rooms are in built using a reentrant NW 50 chamber mounted inside the third room. 
The first evacuation room has two outlet flanges, one is connected to a scroll pump (ANEST IWATA, 500 L/m), and the other to a pressure gauge. 
The second evacuation room is evacuated by a dry screw pump (PFEIFFER Ontool booster pump, 36 L/s), while the third chamber is pumped by a turbo molecular pump (PFEIFFER HiPace300, 300L/s).

\subsection{The SextuPole Ion beam Guide (SPIG)}
A 253 mm long SPIG is mounted parallel to the beam path between the exit hole and the Quadrupole Mass Separator (QMS). 
The SPIG is built from six 1 mm diameter Molybdenum rods. 
The distance between the SPIG and the gas cell exit hole is 1 mm and the diameter of the inner circle is 2 mm.  A pair of RF signals are applied to the SPIG such that adjacent rods receive signals 180$^\circ$ out of phase.  The frequency is fixed at 4.0 MHz and the amplitude can be varied from 0 to 200 Vpp.  A DC offset voltage can be added to the RF signals. 

One concern is whether ions can be transported by SPIG despite the long path (253 mm) in a large pressure gradient. One may expect that all ions are simply stopped by collisions with buffer gas before they reach the high vacuum region. 
In the present case, however, a gas jet assists ions moving inside the SPIG from the high pressure region of the exit hole to the high-vacuum region at the entrance of QMS. 

\subsection{The laser optical system}

\begin{figure}[h]
\begin{center}
\includegraphics*[width= 9.0 cm,height=4.0 cm]{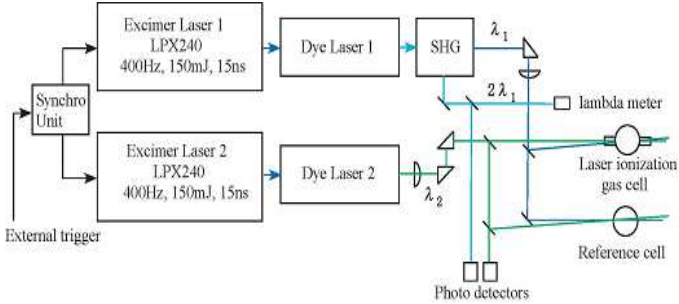}
\end{center}
\caption{Scheme of the laser optics system.}
\label{fig5}
\end{figure}

A laser system for the resonant ionization in off-line setup is shown in Fig.~\ref{fig5}. 
Two XeCl excimer lasers with wavelength $\lambda$ = 308 $nm$ (Lambda Physik LPX240i) pump two dye lasers (Lambda Physik Scanmate, FL3002). Both excimer lasers are synchronized within a precision of a few ns. The maximum repetition rate of the excimer lasers is 400 Hz. We use two-step one- or two-colour schemes for the resonant laser ionization of atoms. The first laser excites atoms from the ground state into an intermediate state followed by a transition into an auto-ionizing state or into the ionization continuum by the second laser. A UV radiation is used for the first step; it can be synthesized by a second harmonic generator (Lambda Physik UV311).  A PC software package has been developed for remotely controlling all laser parameters, adjustment of timing delay, tuning, and scanning of wavelength. In the present prototype system, the laser beams are split into two paths, one path leads to a reference cell where resonance ionization takes place in vacuum for testing the ionization scheme, the other path leads to the laser ionization gas cell.

%
\section{Experimental results and discussion}

\subsection{Pressure distribution and gas jet behavior}

\begin{figure}[h]
\begin{center}
 \includegraphics*[width= 9.0 cm,height=7.0 cm]{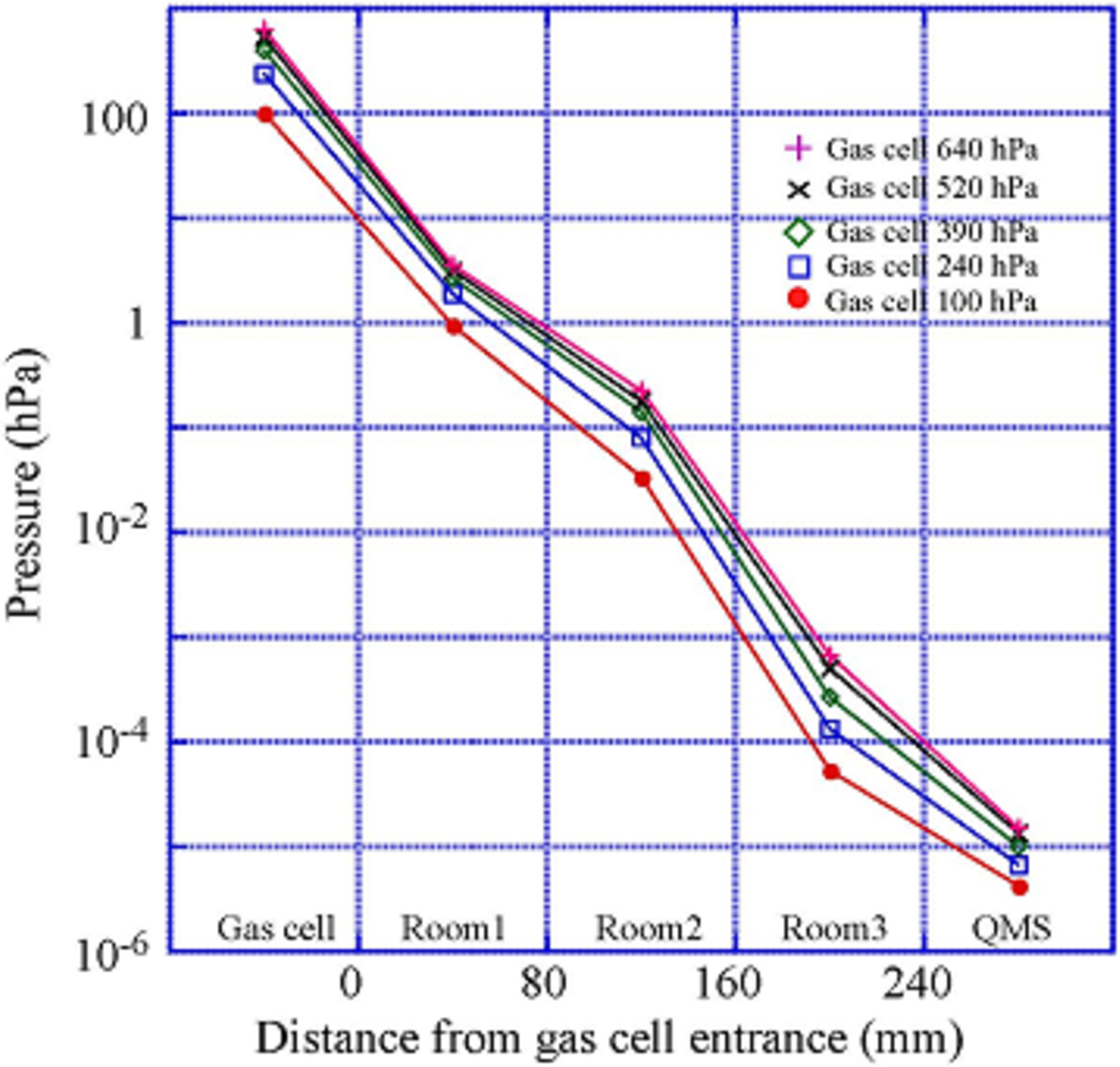}
\end{center}
\caption{Measured pressures at gas cell, each evacuation room in the differential pumping region and QMS. The size of gas cell exit hole diameter is 1 mm.}
\label{fig6}
\end{figure}

The pressure at the individual evacuation rooms were measured by appropriate pressure gauges as shown in Fig.~\ref{fig6}. As described in previous section, in this new differential pumping method, the pressure is gradually decreasing from the gas cell to the QMS. Even with nearly one atmospheric gas cell pressure and 1 mm of exit hole diameter, this differential pumping method is still capable of achieving high vacuum.

By using Eq. (\ref{eqn:b}) and present measured values,  it is possible to see the relative evolution of  the jet component $Q$$_{\rm{jet}}$ which is directly injected into the SPIG.   As $Q$$_{\rm{jet}}$ is a partial flow rate of $Q$, it is given by 

\begin{equation}
Q_{\rm{jet}} = \varepsilon Q ,
\label{eqn:g}
\end{equation}

Then $\varepsilon$ can be written from the relation in Eq. (\ref{eqn:b}) as follows:

\begin{equation}
\varepsilon = 1 - \frac{(P_{1}-P_{2}) C_{1} + P_{1} S_{1}}{Q} .
\label{eqn:h}
\end{equation}

\begin{figure}[h]
\begin{center}
 \includegraphics*[width= 7.0 cm,height=7.0 cm]{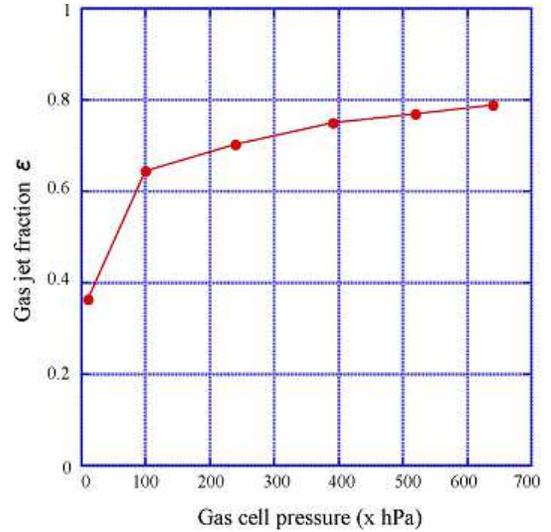}
\end{center}
\caption{Deduced gas jet fraction $\varepsilon$ using measured pressures as function of gas cell pressure.}
\label{fig7}
\end{figure}

In the intermediate vacuum region, the conductance varies with the pressure slightly. 
If we use a fixed value of $C$$_{1}$ = 3.8 $\times$ 10$^{-5}$ m$^{3}$/s which is calculated for the mechanical structure, and substituted $Q$ from Eq.  (\ref{eqn:a}) and $S$$_{1}$ = 0.5 m$^{3}$/min, then the behavior of $\varepsilon$ can be approximated by using experimentally measured pressures $P$$_{1}$ and $P$$_{2}$, as shown in Fig.~\ref{fig7}.

Absolute values of $\varepsilon$ cannot be reliably determined, since the pressure gauges are placed at the wall of chamber in the evacuation rooms, these values differ from that of real beam path.
However it can be presumed that the gas jet component directed into the SPIG increases with the background pressure.  
This means that the divergence of gas jet decreases as the pressure of the first differential pumping region increases \cite{jy1}.
This partial flow rate is regarded as an assisting force gas jet which brings ions to high vacuum during the SPIG transmission.

\subsection{Resonant ionization inside the gas cell, ion extraction and transport to high vacuum}

The resonant laser ionization inside the gas cell, ion extraction and transport to the high-vacuum region via SPIG and QMS have been confirmed. So far we have tested ionization inside the gas cell for stable isotopes of the following elements: Ni, Fe, Cu, Co, Pd, Sn, Ti, Nb and In.  The ionization was performed by one colour or two colours two-step excitation scheme \cite{ris2012}. 

\begin{figure}[h]
\begin{center}
 \includegraphics*[width= 6.0 cm,height=12.0 cm]{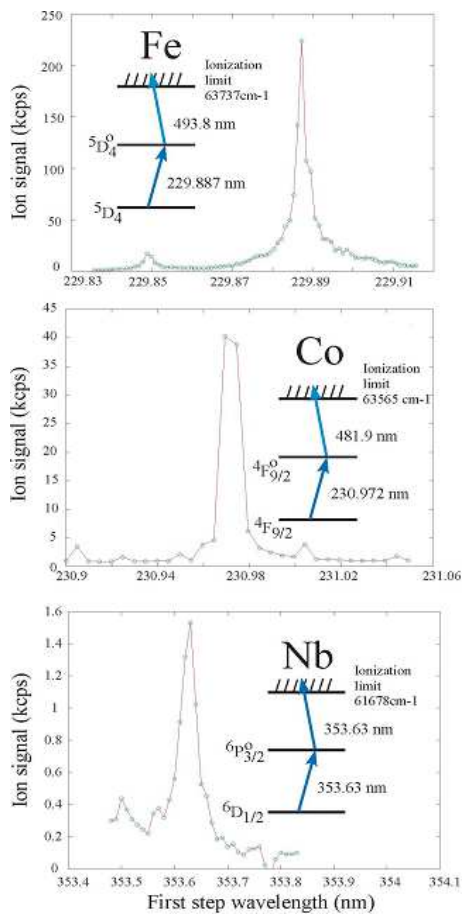}
\end{center}
\caption{Scans of the first step laser wavelength for resonance ionization of Fe, Co and Nb. The second step laser was tuned to wavelengths for  the auto-ionizing state in case of Fe and Co.}
\label{fig8}
\end{figure}

Atoms to be photo-ionized were produced by  either evaporation of a filament or YAG laser ablation inside the cell.  The gas cell pressure was adjusted from 100 to 700 hPa argon. First, evaporated atoms are released in the middle region of the gas cell, then they move via gas flow to the laser ionization region just before the gas cell exit. Two dye laser beams are focused on a quartz view port and irradiate the flowing atoms for ionization.
The laser repetition rate was typically set between 10-50 Hz during the experiment, though this value can be increased to a maximum of 400 Hz. 

Since the laser wavelength is tuned for an element-specific ionization scheme which is determined by an ionizing atomic transition,
we can identify the ion of interest with an element separation by laser in addition to a mass separation by QMS.  
After ionization inside the gas cell, photo-ionized ions are injected into the SPIG with a gas jet, and pass through the QMS entrance with 10 mm aperture, providing mass selection prior to CEM. Figure~\ref{fig8} shows example detected ion signals for Fe, Co and Nb in scans of the first step excitation laser wavelength.  Ion detection rate depends on the status of filament or ablation target; however, normally more than 10$^{4}$ counts per second (cps) are observed with 50 Hz laser repetition rate.   Saturation of ionization efficiency was observed in each element, typically around 100 $\mu$J/pulse in the first transition, 1 mJ/pulse in the auto-ionizing transition.

\subsection{The investigation of SPIG performance} 
The investigation of SPIG performance was carried out by both simulation and experiment.  
In simulation, the trajectories for individual ions were microscopically traced step by step using a Monte Carlo technique until the ions either collide with SPIG rods or reach the end of the SPIG.
The effects of the RF + DC electric fields and gas flow affecting the motion of ions were calculated separately.
The Navier-Stokes flow calculation provides flow velocity, temperature and density for the gas. 
The calculation used a mesh with dimensions of 10-100 $\mu$m, limited by the computing memory and reasonable time of computation. The mean free path of ions was determined by the density, while the velocity of the buffer gas atoms was calculated by a vector summation of the flow velocity and the thermal motion of gas atoms following the Maxwell-Boltzmann distribution. 
Electric field was calculated by SIMION \cite{simion}. Between collisions, the  motion of the ions under the influence of the RF+DC electric fields was determined by means of  a 4$^{\rm{th}}$ order Runge-Kutta method. Collisions kinematics were calculated by the rigid sphere model. While there is no way to exactly reproduce the experimental result by simulation without knowing the real pressure distribution inside the SPIG, it is possible to understand the behavior of ion transmission by calculating  pressure distribution using measured pressures as boundary conditions.  In this work, the tested  boundary conditions were as follows:  500 hPa in the gas cell, 10 hPa in the first differential pumping room, 1 hPa in the second differential pumping room, and 10$^{-3}$ hPa in the third differential pumping room.
A continuous pressure gradient, based on these boundary conditions, was utilized in simulations.

\begin{figure}[h]
\begin{center}
 \includegraphics*[width=8cm]{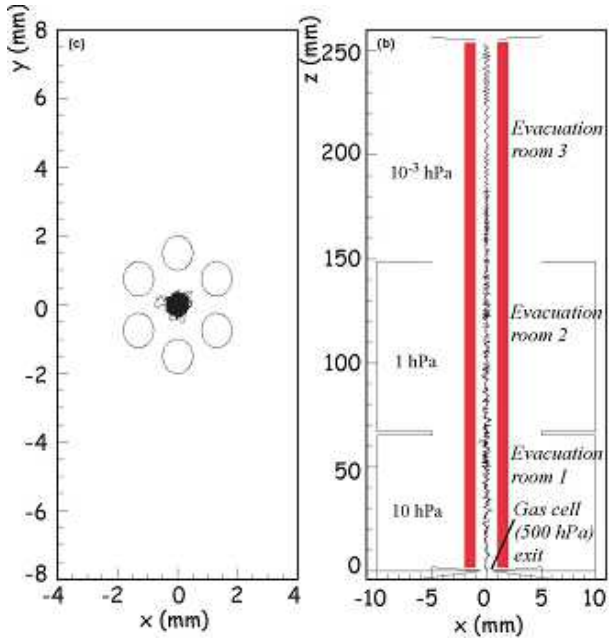}
\end{center}
\caption{Monte Carlo simulation of an ion's motion in the SPIG. An ion (${m}$ = 56) starting at the gas cell exit hole follows the gas flow, along the z axis, of the assisting force gas jet while trapped in the radial potential well of the SPIG RF field.}
\label{fig9}
\end{figure}

In this simulation, the initial ion distribution was randomly populated, using a guassian random number generator by width of 0.5 mm, on a surface filling the opening of the exit hole. Space charge effects and ion loss due to interaction between impurities inside buffer gas were not considered. A typical trajectory for an Fe$^+$ ion moving through the SPIG in the simulation is shown in Fig.~\ref{fig9}; the simulation used SPIG RF signals of 100 V$_{\rm{pp}}$ at 4.0 MHz. 
From this figure, we can see that the ion can be transmitted through the SPIG, even in the case of relatively dense pressure gradient. 

\begin{figure}[h]
\begin{center}
 \includegraphics*[width= 7.0 cm,height=7.0 cm]{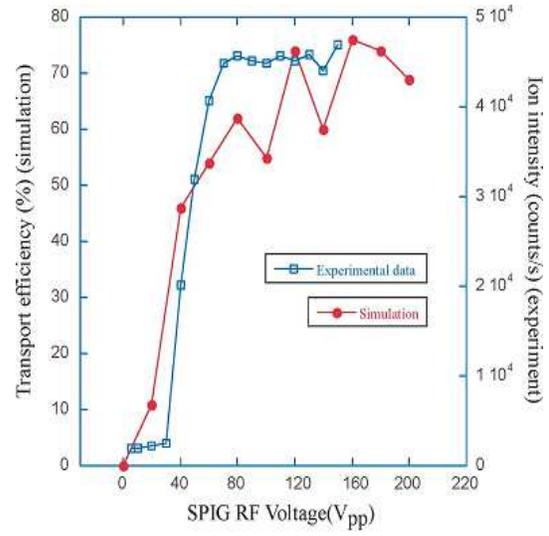}
\end{center}
\caption{Simulated and measured SPIG transmission of Fe$^+$ as functions of RF voltage applied to SPIG.  Simulations followed ions to the end of the SPIG, while measured data required ions to pass through a QMS set to select for $\frac{m}{q} = 56$}
\label{fig10}
\end{figure}

The simulated transmission efficiency as a function of RF voltage is shown in Fig.~\ref{fig10} for the pressure distribution described above. The graph also  includes experimental data, measured with 500 hPa argon in the gas cell. The simulated efficiency scale is given along the left vertical axis, while the right vertical axis denotes experimentally measured ion intensity.  The simulated efficiency is based on initial ion distribution of 100 ions with atomic mass of ${m}$ = 56.  
 The experimental data was measured by operating the QMS with RF and DC potentials appropriate for rejecting all ions with $m/q \ne 56$ without DC potential difference between SPIG and QMS.
The trend in the experiment and calculated curves agree well.  A saturation voltage, where a radial focusing force is enough to confine ions inside the SPIG, was found near 80 V$_{\rm{pp}}$, in good agreement with the simulation.  

\begin{figure}[h]
\begin{center}
 \includegraphics*[width=8cm]{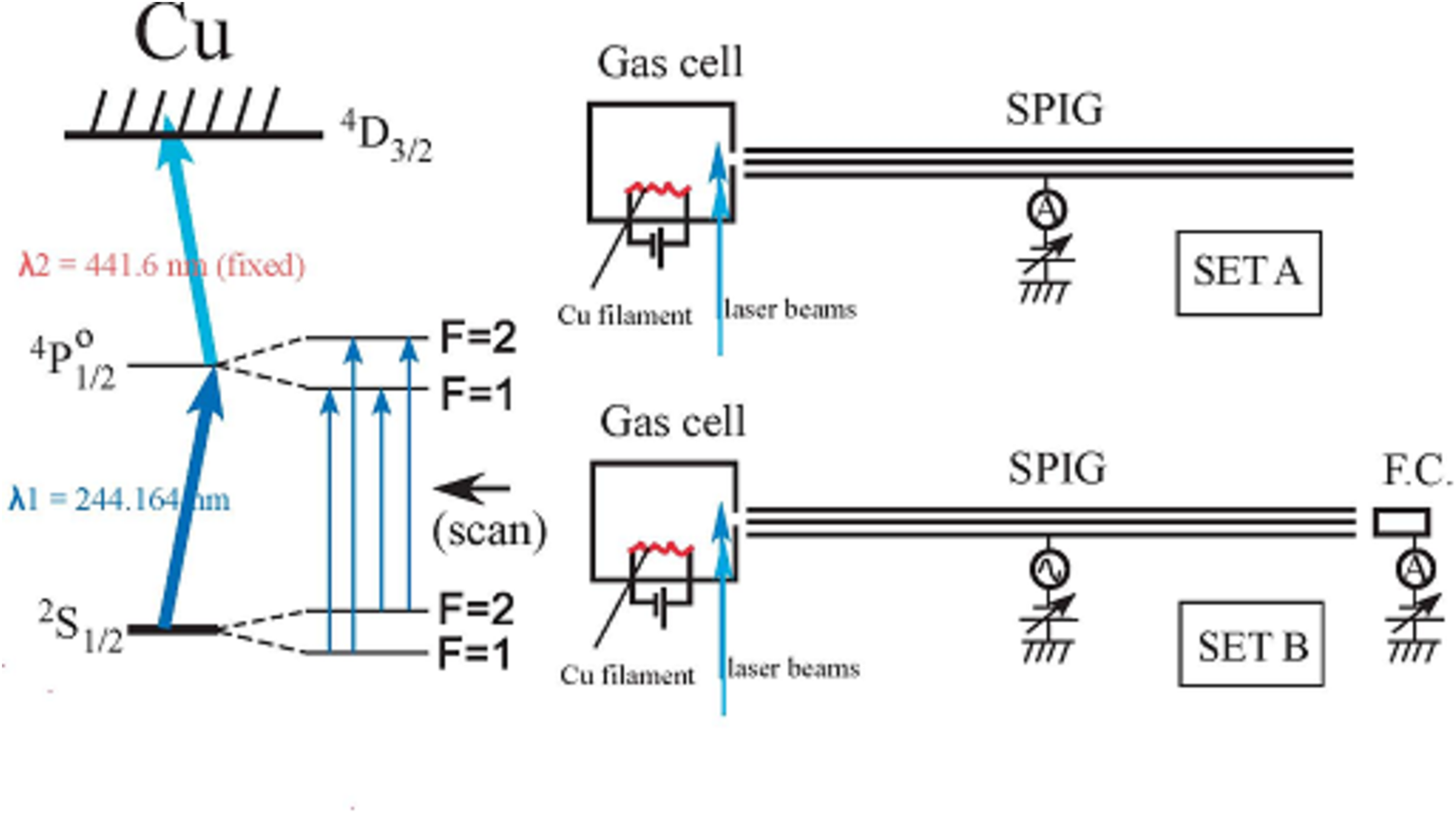}
\end{center}
\caption{Layout of the apparatus used for the measurement of the SPIG transmission efficiency.
SET A was used for the measurement of the total ion current extracted from the gas cell. The ions were collected on on the SPIG rods, without RF signal.  While SET B was used to measure the ion current transmitted to the end of SPIG operated with RF signals of 100 V$_{\rm{pp}}$ at 4.0 MHz. Atomic level diagram of Cu indicated on the left side is the ionization scheme used in the present experiment.}
\label{fig10n}
\end{figure}

\begin{figure}[h]
\begin{center}
 \includegraphics*[width=8cm]{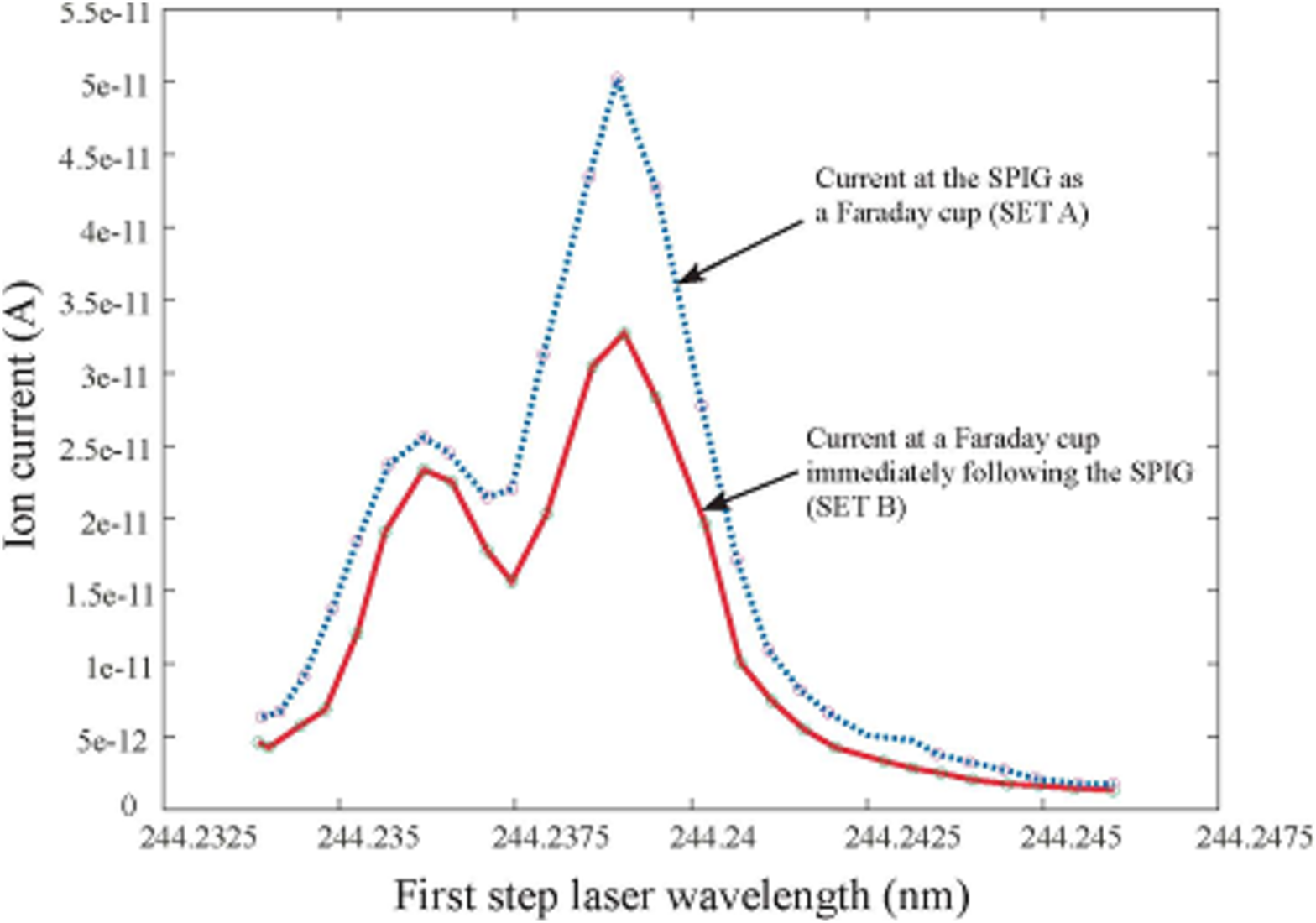}
\end{center}
\caption{Ion current as function of first-step laser wavelength, measured during laser resonance ionization of Cu.  The dashed line is current measured on the SPIG rods, without RF signal applied (SET A).  The solid is current measured at a Faraday cup immediately following the SPIG, when the SPIG was operated with RF signals of 100 V$_{\rm{pp}}$ at 4.0 MHz (SET B).  From the integrated value of the two curves, the absolute SPIG transimission efficiency was 60-80\%.}
\label{fig11}
\end{figure}

Separately, it was possible to determine the absolute transmission efficiency for the SPIG. The experimental setups and the used ionization scheme for Cu are shown in Fig.~\ref{fig10n}. The efficiency is defined as the ratio of the ion current operated with RF signals of 100 V$_{pp}$ at 4.0 MHz collected on a Faraday cup placed after the SPIG (SET B in Fig.~\ref{fig10n}) to the ion current collected on the SPIG rods without RF signal applied (SET A in Fig.~\ref{fig10n}). This was done during a test of resonant laser ionization for stable Cu.  Comparing  from the integrated value of two ionization curves, shown in Fig.~\ref{fig11}, an absolute SPIG transimission efficiency was 60-80\% with a gas cell pressure of 500 hPa. 

\subsection{Time profiles of photo-ionized ions as a probe of SPIG transit time and gas cell extraction time}

The time profile of ions extracted from the gas cell provides an important probe to understand the behavior of ions moving inside the gas cell and the SPIG.  
In this work, we have examined the time profile in two trigger modes.  Using the ionization laser pulse as a start signal probes transmission time  through the SPIG, while using a YAG laser pulse as a start signal allows the use of ablated material as a probe of extraction time from the gas cell. Ion detection at the CEM provides the stop signal in both cases.

\subsubsection{Transport of ions through the SPIG}

\begin{figure}[h]
\begin{center}
 \includegraphics*[width=8cm]{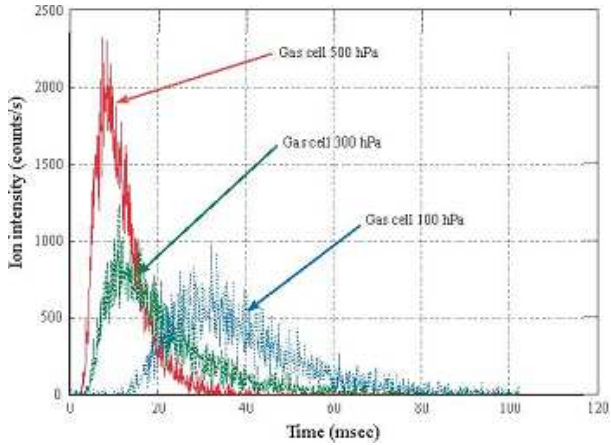}
\end{center}
\caption{The measured extraction time profile from the ionization region to the CEM ion detector.}
\label{fig12}
\end{figure}

It is necessary to know the time required for ion transport through the differential pumping region.  Due to the high pressure distribution in the SPIG, it is not obvious that the transport will be sufficiently fast.  To investigate this effect, the ionization lasers were operated at 10 Hz and serve as a start signal for the timing measurement. Fe vapor atoms were continuously released from a hot filament inside the gas cell. They were resonantly ionized and finally measured at the CEM behind the QMS, with the QMS set to selectively transport ions with $m/q = 56$.  The time profile results for Fe$^+$ using various gas cell pressures are shown in Fig.~\ref{fig12}.  The signal represents a convolution of three components: transport from the ionization region to the gas cell exit hole, transport through the SPIG and transport through the QMS.  We can neglect the transport time through the QMS, as it is located in high-vacuum and transport should occur in the order of tens of microseconds.

These results show that the transport time for ions becomes faster with increasing gas cell pressure. For the first order, the extraction time in the gas cell does not depend on the pressure, implying that this phenomena is caused by the reduction of the SPIG transmission time. This can be understood from the shape of gas jet. When the pressure in the gas cell increases, the background pressure in the first differential pumping room is also increased and the divergence angle of the gas jet becomes small as described in section 3.1 and \cite{jy1}.  This causes the effective length of gas jet inside the SPIG to increase.  As a result, successive collisions force ions towards the high-vacuum region. This reduces transmission time for ions in the SPIG.  On the other hand, once the gas jet disappears, ions are no longer effectively pushed, which leads to longer transmission times at lower gas cell pressures. 

\begin{figure}[h]
\begin{center}
 \includegraphics*[width=12.1cm,height=9.1 cm]{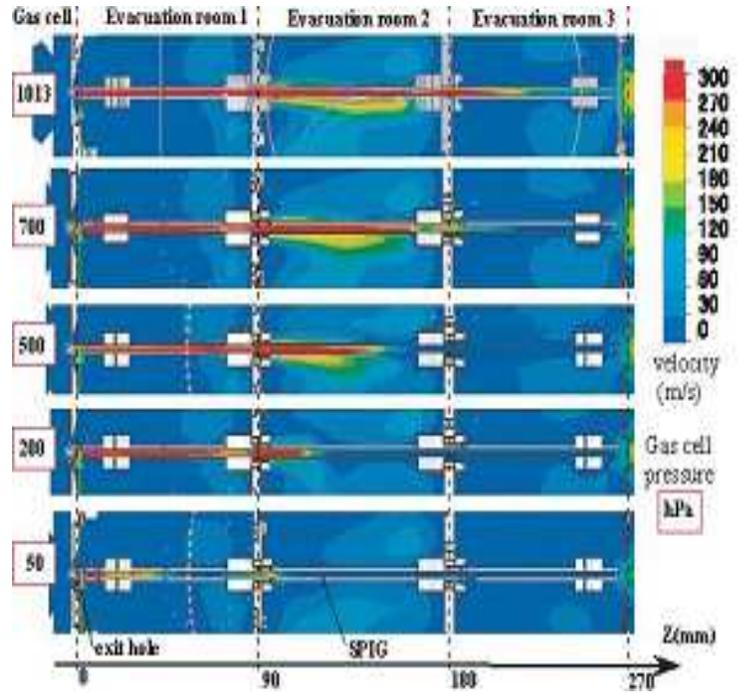}
\end{center}
\caption{Simulated gas velocity distribution at different pressure conditions in individual room. The colour scale covers the range 0 to 300 $m/s$. Colour online.}
\label{fig13}
\end{figure}

Figure~\ref{fig13} shows the flow simulation result revealing the gas velocity distribution for various gas cell pressures at cutting plane of the  central axis of the SPIG.  The boundary conditions for gas cell pressure used in the simulation are shown at the left side in the figure.  
It can be seen that the simulated  gas jet length in terms of the gas velocity over 300 $\rm{m/s}$ (red colour) becomes longer as the gas cell pressure  increases.
This gas jet length indicates the effective range of the assisting force gas jet, such the higher velocity gas jet brings ions to the vacuum region quickly. 

We can conclude from the qualitative agreement between simulation and experiment that the jet structure determines the transit time of ions inside the SPIG.  This assisting force gas jet effect can be utilized to provide effective transmission of ions into the high-vacuum region despite the high background pressure, extending the permissible upper pressure limit in the gas cell.

\subsubsection{Extraction of atoms from the gas cell by gas flow}

\begin{figure}[h]
\begin{center}
 \includegraphics*[width=8cm]{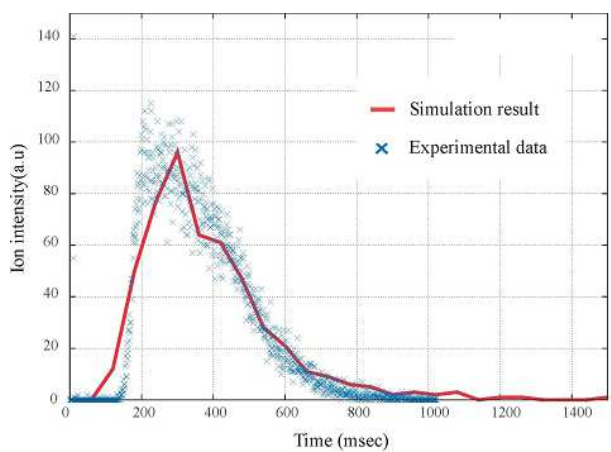}
\end{center}
\caption{The time profile for atom's released by YAG laser ablation in the gas cell, photo-ionized at the exit nozzle and measured at CEM.  Good agreement can be seen between measured (points) and simulated (line) results.}
\label{fig14}
\end{figure}

A laser ablation target was installed in the gas cell, approximately 60 mm from the gas cell exit nozzle.  This allows for the pulsed production of atomic ensemble by YAG laser.  
By operating the YAG laser at a repetition rate of 1 Hz, while the repetition rate of ionization lasers is 100 Hz, it is possible to study the transport time of atoms in the gas cell.  Figure~\ref{fig14} compares simulated and measured time profiles for stable Fe, using the ablation pulse of the YAG laser as the start signal, with gas cell pressure of 100 hPa.

When the wavelength of the ionization laser was off-resonance, no ion signal was observed with QMS selectively transmitting ions with $m/q = 56$.  This indicates that most Fe$^+$ ions produced by YAG laser ablation are neutralized in the gas cell.  The time-scale is an order of magnitude larger than that for SPIG transmission time shown in Fig.~\ref{fig12}.  
Most of the extraction time is spent transporting ions from a region near the ablation target to the exit of the cell.
The broad time-structure in Fig. \ref{fig14} can be explained by the initial distribution of atoms in the ablation plume.

The simulated results were calculated using macroscopic motion combined with gas flow calculations.  The gas flow calculation was performed using a precise 3D model of the gas cell which took the full geometrical complexity into account.  The gas flow calculation provided flow velocities at each point on a 3D mesh.   A random number generator was used to create an initial distribution of the atomic plume in a region near the ablation target surface.  Trajectories were then calculated from the gas flow, in combination with diffusion of the atomic plume, by means of a fixed time-step ray-tracing algorithm until collision with the chamber wall or transport to the exit region. Good agreement is seen between the two results. 

The experimental data include transport time through the SPIG and QMS, while the simulation does not include these transmission times as it ends when ions reach the gas cell exit hole. This makes for a discrepancy between experiment and simulation, because the rise time from simulations is artificially fast from lacking transmission times trough the SPIG and QMS.

\section{Summary and outlook}
We have shown proof of principle studies for a resonant laser ionization gas cell that utilizes a novel differential pumping method, based on the use of a few small scroll pumps.  In terms of maximum allowable gas cell pressures, the new method was shown to be comparable or superior to conventional differential pumping systems based on large roots or screw pumps configuration. Furthermore the size of the differential pumping system has been drastically reduced. It will open a new possibility for breaking through the limitations on gas cell pressure and exit hole size, while allowing for a very compact assembly.

Off-line tests of resonant laser ionization for stable Co, Cu, Fe, Ni, Ti, Pd, Nb, Sn and In inside the gas cell, along with extraction of these ions to high-vacuum, was successful.  The ability to photo-ionize atomic vapors and transport the ions through a long SPIG was confirmed. The fast ion transport can be attributed to the affect of an assisting force gas jet.
This gas jet is an effect of allowing a high pressure in the first differential pumping room, which results in a narrow, collimated gas jet. This narrow, collimated gas jet forces ions through the SPIG quickly.  As another spin-off, this jet structure is brought for ideal environment for laser spectroscopy\cite{ls4}\cite{ls3}\cite{list2}\cite{jy0}. The first result for the fundamental study of gas-jet laser spectroscopy in the present setup can be found in \cite{tka}. 

Simulation of atomic transport in the gas cell and ion transport through the SPIG was experimentally verified.  By using the ionizing laser pulse as a start signal, it was possible to study the transit time through the SPIG.  The result showed an inverse relationship between gas cell pressure and SPIG transit time.  This is a powerful verification of the existence of an assisting force gas jet.  Alternately, the use of a YAG laser to produce pulsed atomic vapor plumes allowed the study of transport inside the gas cell. 
The excellent agreement of these measurements with simulations based on gas flow calculations allows us to confidently estimate the expected yield of short-lived nuclei for newly designed PALIS gas cell.

\begin{figure}[h]
\begin{center}
 \includegraphics*[width=8cm]{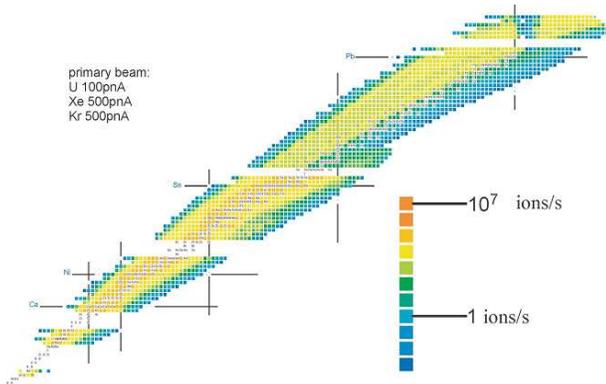}
\end{center}
\caption{The expected yield of available elements by laser ionization for PALIS on-line experiments. The estimation includes efficiencies attributed to individual processes, such as stopping in the gas cell, laser ionization and decay loss for short-lived nuclei.}
\label{fig15}
\end{figure}

The miniaturization of the differential pumping system, along with the ability to use higher gas pressures, will enable installation of the parasitic laser ion source (PALIS) in the limited space of the BigRIPS fragment separator.  As the first phase, PALIS will be installed in the vicinity of F2 focal plane in the BigRIPS fragment separator. Detailed design work for the gas cell, differential pumping system and the low energy RI-beam transport line is currently in progress. 
The expected PALIS yield, based on extraction time from gas cell and ability to photo-ionize, is shown in Fig.~\ref{fig15}.  
The availability of PALIS system will result in a many-fold increase in the available beam time. Furthermore, in combination with narrow gas jet structure, the laser spectroscopy inside the gas jet will be a vital tool to explore the study of the ground state property for wide range of rare exotic nuclei provided by BigRIPS.

\section{Acknowledgments}
We wish to thank RIKEN Nishina Center for Accelerator-Based Science for financial support of this research.
We would like to thank Dr. M.~Wakasugi of RIKEN,  Dr. M.~Koizumi and Dr. M.~Ohba of JAEA and Dr. T.~Mitsugashira of IRCNMS for providing us their laser components. Also we would like to express our appreciation of Mr. H.~Imamura of Hakuto Co.,Ltd, Mr. T. Kimura of Coherent Japan, Ms. W. Kina of Indeco, INC. and Dr. T. Kambara and Dr. T.~Kobayashi of RIKEN for their kind support. We are grateful to M.~Huyse, P.~Van Duppen, Yu.~Kudryavtsev, R.~Ferrer and P.~Van~den~Bergh of K.U.Leuven for their support and fruitful discussions during development.

\end{document}